\pdfoutput=1
\documentclass[12pt,a4paper]{article}

\usepackage{ifthen} 
\newboolean{pdflatex}
\setboolean{pdflatex}{true} 

\newboolean{articletitles}
\setboolean{articletitles}{true} 

\newboolean{uprightparticles}
\setboolean{uprightparticles}{false} 

\def\paperauthors{LHCb collaboration} 
\def\paperasciititle{Search for the lepton flavor violating decay Bs to phi mu tau} 
\def\papertitle{Search for the lepton-flavor violating decay $B^0_s\to\phi\mu^\pm\tau^\mp$} 
\def\paperkeywords{{High Energy Physics}, {LHCb}} 
\def\papercopyright{\the\year\ CERN for the benefit of the LHCb collaboration} 
\def\paperlicence{CC BY 4.0 licence}
\def\paperlicenceurl{https://creativecommons.org/licenses/by/4.0/}

\usepackage[top=1in, bottom=1.25in, left=1in, right=1in]{geometry}

\usepackage{microtype}
\usepackage{lineno}  
\usepackage{xspace} 
\usepackage{caption} 

\usepackage{graphicx}  
\usepackage{color}
\usepackage{colortbl}
\graphicspath{{./figs/}} 

\usepackage{amsmath} 
\usepackage{amssymb}
\usepackage{amsfonts}
\usepackage{upgreek} 

\newcommand*\patchAmsMathEnvironmentForLineno[1]{%
\expandafter\let\csname old#1\expandafter\endcsname\csname #1\endcsname
\expandafter\let\csname oldend#1\expandafter\endcsname\csname
end#1\endcsname
 \renewenvironment{#1}%
   {\linenomath\csname old#1\endcsname}%
   {\csname oldend#1\endcsname\endlinenomath}%
}
\newcommand*\patchBothAmsMathEnvironmentsForLineno[1]{%
  \patchAmsMathEnvironmentForLineno{#1}%
  \patchAmsMathEnvironmentForLineno{#1*}%
}
\AtBeginDocument{%
\patchBothAmsMathEnvironmentsForLineno{equation}%
\patchBothAmsMathEnvironmentsForLineno{align}%
\patchBothAmsMathEnvironmentsForLineno{flalign}%
\patchBothAmsMathEnvironmentsForLineno{alignat}%
\patchBothAmsMathEnvironmentsForLineno{gather}%
\patchBothAmsMathEnvironmentsForLineno{multline}%
\patchBothAmsMathEnvironmentsForLineno{eqnarray}%
}

\usepackage{hyperxmp}

\usepackage[pdftex,
            pdfauthor={\paperauthors},
            pdftitle={\paperasciititle},
            pdfkeywords={\paperkeywords},
            pdfcopyright={Copyright (C) \papercopyright},
            pdflicenseurl={\paperlicenceurl}]{hyperref}

\usepackage[colorinlistoftodos,textsize=scriptsize]{todonotes}

\usepackage[bottom,flushmargin,hang,multiple]{footmisc}

\usepackage[all]{hypcap} 

\usepackage{xspace} 
\usepackage{upgreek}

\def\lhcb   {\mbox{LHCb}\xspace}

\def\MagUp {\mbox{\em Mag\kern -0.05em Up}\xspace}

\ifthenelse{\boolean{uprightparticles}}%
{

 \def\Pmu         {\ensuremath{\upmu}\xspace}                 
 \def\Pnu         {\ensuremath{\upnu}\xspace}                 
                  
 \def\Ppi         {\ensuremath{\uppi}\xspace}                 
                  
 \def\Prho        {\ensuremath{\uprho}\xspace}                 
                  
 \def\Ptau        {\ensuremath{\uptau}\xspace}

 \def\Ppsi        {\ensuremath{\uppsi}\xspace}

 \def\PDelta      {\ensuremath{\Delta}\xspace}                 
 \def\PXi         {\ensuremath{\Xi}\xspace}                 
 \def\PLambda     {\ensuremath{\Lambda}\xspace}                 
 \def\PSigma      {\ensuremath{\Sigma}\xspace}                 
 \def\POmega      {\ensuremath{\Omega}\xspace}                 
 \def\PUpsilon    {\ensuremath{\Upsilon}\xspace}
 \let\oldPi\Pi
 \def\PPi         {\ensuremath{\oldPi}\xspace}

 \def\PB      {\ensuremath{\mathrm{B}}\xspace}                 
                  
 \def\PD      {\ensuremath{\mathrm{D}}\xspace}

 \def\PJ      {\ensuremath{\mathrm{J}}\xspace}                 
 \def\PK      {\ensuremath{\mathrm{K}}\xspace}

 \def\Pb      {\ensuremath{\mathrm{b}}\xspace}                 
 \def\Pc      {\ensuremath{\mathrm{c}}\xspace}

 \def\Pi      {\ensuremath{\mathrm{i}}\xspace}

 \def\Ps      {\ensuremath{\mathrm{s}}\xspace}

 \def\thebaroffset{0.0em}
}
{

 \def\Pmu         {\ensuremath{\mu}\xspace}                 
 \def\Pnu         {\ensuremath{\nu}\xspace}                 
                  
 \def\Ppi         {\ensuremath{\pi}\xspace}                 
                  
 \def\Prho        {\ensuremath{\rho}\xspace}                 
                  
 \def\Ptau        {\ensuremath{\tau}\xspace}

 \def\Ppsi        {\ensuremath{\psi}\xspace}                 
                  
 \mathchardef\PDelta="7101
 \mathchardef\PXi="7104
 \mathchardef\PLambda="7103
 \mathchardef\PSigma="7106
 \mathchardef\POmega="710A
 \mathchardef\PUpsilon="7107
 \mathchardef\PPi="7105
                  
 \def\PB      {\ensuremath{B}\xspace}                 
                  
 \def\PD      {\ensuremath{D}\xspace}

 \def\PJ      {\ensuremath{J}\xspace}                 
 \def\PK      {\ensuremath{K}\xspace}

 \def\Pb      {\ensuremath{b}\xspace}                 
 \def\Pc      {\ensuremath{c}\xspace}

 \def\Pi      {\ensuremath{i}\xspace}

 \def\Ps      {\ensuremath{s}\xspace}

 \def\thebaroffset{0.18em}
}
\newcommand{\offsetoverline}[2][\thebaroffset]{\kern #1\overline{\kern -#1 #2}}%

\makeatletter
\ifcase \@ptsize \relax
  \newcommand{\miniscule}{\@setfontsize\miniscule{4}{5}}
\or
  \newcommand{\miniscule}{\@setfontsize\miniscule{5}{6}}
\or
  \newcommand{\miniscule}{\@setfontsize\miniscule{5}{6}}
\fi
\makeatother

\DeclareRobustCommand{\optbar}[1]{\shortstack{{\miniscule (\rule[.5ex]{1.25em}{.18mm})}
  \\ [-.7ex] $#1$}}

\def\mup        {{\ensuremath{\Pmu^+}}\xspace}
\def\mun        {{\ensuremath{\Pmu^-}}\xspace} 
\def\mupm       {{\ensuremath{\Pmu^\pm}}\xspace}

\def\taup       {{\ensuremath{\Ptau^+}}\xspace}
\def\taum       {{\ensuremath{\Ptau^-}}\xspace}

\def\taump      {{\ensuremath{\Ptau^\mp}}\xspace}

\def\ellm       {{\ensuremath{\ell^-}}\xspace}

\def\neu        {{\ensuremath{\Pnu}}\xspace}
\def\neub       {{\ensuremath{\overline{\Pnu}}}\xspace}
\def\neut       {{\ensuremath{\neu_\tau}}\xspace}

\def\neulb      {{\ensuremath{\neub_\ell}}\xspace}

\def\squark    {{\ensuremath{\Ps}}\xspace}

\def\cquark    {{\ensuremath{\Pc}}\xspace}

\def\bquark    {{\ensuremath{\Pb}}\xspace}

\def\pion   {{\ensuremath{\Ppi}}\xspace}
\def\piz    {{\ensuremath{\pion^0}}\xspace}
\def\pip    {{\ensuremath{\pion^+}}\xspace}
\def\pim    {{\ensuremath{\pion^-}}\xspace}

\def\rhomeson {{\ensuremath{\Prho}}\xspace}
\def\rhoz     {{\ensuremath{\rhomeson^0}}\xspace}

\def\kaon    {{\ensuremath{\PK}}\xspace}

\def\KorKbar {\kern \thebaroffset\optbar{\kern -\thebaroffset \PK}{}\xspace}

\def\Kp      {{\ensuremath{\kaon^+}}\xspace}
\def\Km      {{\ensuremath{\kaon^-}}\xspace}

\def\KS      {{\ensuremath{\kaon^0_{\mathrm{S}}}}\xspace}

\def\KL      {{\ensuremath{\kaon^0_{\mathrm{L}}}}\xspace}

\def\Dbar    {{\ensuremath{\offsetoverline{\PD}}}\xspace}
\def\D       {{\ensuremath{\PD}}\xspace}

\def\DorDbar {\kern \thebaroffset\optbar{\kern -\thebaroffset \PD}\xspace}

\def\Dzb     {{\ensuremath{\Dbar{}^0}}\xspace}
\def\Dp      {{\ensuremath{\D^+}}\xspace}
\def\Dm      {{\ensuremath{\D^-}}\xspace}

\def\DpDm    {\ensuremath{\Dp {\kern -0.16em \Dm}}\xspace}

\def\Dsm     {{\ensuremath{\D^-_\squark}}\xspace}

\def\B       {{\ensuremath{\PB}}\xspace}
\def\Bbar    {{\ensuremath{\offsetoverline{\PB}}}\xspace}

\def\BorBbar {\kern \thebaroffset\optbar{\kern -\thebaroffset \PB}\xspace}

\def\Bd      {{\ensuremath{\B^0}}\xspace}

\def\BdorBdbar {\kern \thebaroffset\optbar{\kern -\thebaroffset \Bd}\xspace}
\def\Bu      {{\ensuremath{\B^+}}\xspace}

\def\Bs      {{\ensuremath{\B^0_\squark}}\xspace}
\def\Bsb     {{\ensuremath{\Bbar{}^0_\squark}}\xspace}
\def\BsorBsbar {\kern \thebaroffset\optbar{\kern -\thebaroffset \Bs}\xspace}

\def\jpsi     {{\ensuremath{{\PJ\mskip -3mu/\mskip -2mu\Ppsi}}}\xspace}
\def\psitwos  {{\ensuremath{\Ppsi{(2S)}}}\xspace}

\def\Y#1S{\ensuremath{\PUpsilon{(#1S)}}\xspace}

\def\LorLbar     {\kern \thebaroffset\optbar{\kern -\thebaroffset \PLambda}\xspace}

\def\BF         {{\ensuremath{\mathcal{B}}}\xspace}

\newcommand{\decay}[2]{\ensuremath{#1\!\to #2}\xspace} 

\def\to                 {\ensuremath{\rightarrow}\xspace}

\def\AT#1     {\ensuremath{A_{\mathrm{T}}^{#1}}\xspace}           

\def\C#1      {\ensuremath{\mathcal{C}_{#1}}\xspace}                       
\def\Cp#1     {\ensuremath{\mathcal{C}_{#1}^{'}}\xspace}                    
\def\Ceff#1   {\ensuremath{\mathcal{C}_{#1}^{\mathrm{(eff)}}}\xspace}        
\def\Cpeff#1  {\ensuremath{\mathcal{C}_{#1}^{'\mathrm{(eff)}}}\xspace}       
\def\Ope#1    {\ensuremath{\mathcal{O}_{#1}}\xspace}                       
\def\Opep#1   {\ensuremath{\mathcal{O}_{#1}^{'}}\xspace}                    

\newcommand{\nospaceunit}[1]{\ensuremath{\text{#1}}}       
\newcommand{\aunit}[1]{\ensuremath{\text{\,#1}}}       

\newcommand{\tev}{\aunit{Te\kern -0.1em V}\xspace}
\newcommand{\gev}{\aunit{Ge\kern -0.1em V}\xspace}
\newcommand{\mev}{\aunit{Me\kern -0.1em V}\xspace}
\newcommand{\kev}{\aunit{ke\kern -0.1em V}\xspace}
\newcommand{\ev}{\aunit{e\kern -0.1em V}\xspace}
 
\newcommand{\mevc}{\ensuremath{\aunit{Me\kern -0.1em V\!/}c}\xspace}
\newcommand{\gevc}{\ensuremath{\aunit{Ge\kern -0.1em V\!/}c}\xspace}
\newcommand{\mevcc}{\ensuremath{\aunit{Me\kern -0.1em V\!/}c^2}\xspace}
\newcommand{\gevcc}{\ensuremath{\aunit{Ge\kern -0.1em V\!/}c^2}\xspace}

\def\mum  {\ensuremath{\,\upmu\nospaceunit{m}}\xspace}

\def\fb   {\ensuremath{\aunit{fb}}\xspace}
\def\invfb   {\ensuremath{\fb^{-1}}\xspace}

\def\gsim{{~\raise.15em\hbox{$>$}\kern-.85em
          \lower.35em\hbox{$\sim$}~}\xspace}
\def\lsim{{~\raise.15em\hbox{$<$}\kern-.85em
          \lower.35em\hbox{$\sim$}~}\xspace}

\def\pt         {\ensuremath{p_{\mathrm{T}}}\xspace}

\def\ptot       {\ensuremath{p}\xspace}

\def\evtgen     {\mbox{\textsc{EvtGen}}\xspace}

\def\geant      {\mbox{\textsc{Geant4}}\xspace}

\def\pythia     {\mbox{\textsc{Pythia}}\xspace}

\def\tell1  {TELL1\xspace}
\def\ukl1   {UKL1\xspace}

\newcommand{\lhcborcid}[1]{\href{https://orcid.org/#1}{\hspace*{0.1em}\raisebox{-0.45ex}{\includegraphics[width=1em]{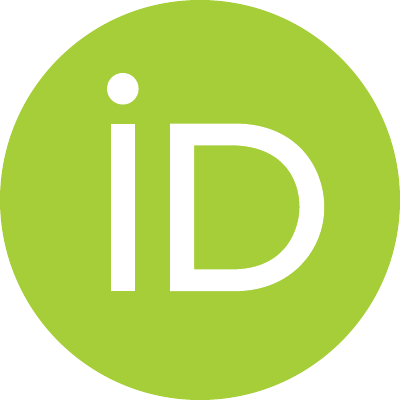}}}}

\usepackage{cite} 
\usepackage{mciteplus}
\usepackage{subcaption}

\usepackage[separate-uncertainty=true]{siunitx}
\DeclareSIUnit\evolt{e\kern -0.1em V}
\usepackage{physics}
\usepackage{booktabs}
\usepackage[capitalize]{cleveref}
\usepackage{multirow}

\def\bsphimutau{\decay{\Bs}{\phi\mup\taum}}

\def\kkmu{\ensuremath{\Kp\Km\mup}\xspace}
\def\thpi{\ensuremath{3\pi}\xspace}
\def\thpiz{\ensuremath{3\pi\piz}\xspace}
\def\mfit{\ensuremath{m_{\text{fit}}}\xspace}

\usepackage{accents}
\newlength{\dhatheight}
\newcommand{\doublehat}[1]{%
    \settoheight{\dhatheight}{\ensuremath{\hat{#1}}}%
    \addtolength{\dhatheight}{-1.0ex}%
    \hat{\vphantom{\rule{1pt}{\dhatheight}}%
    \smash{\hat{#1}}}}

\usepackage{longtable} 

\begin{document}

\renewcommand{\thefootnote}{\fnsymbol{footnote}}
\setcounter{footnote}{1}


\begin{titlepage}
\pagenumbering{roman}

\vspace*{-1.5cm}
\centerline{\large EUROPEAN ORGANIZATION FOR NUCLEAR RESEARCH (CERN)}
\vspace*{1.5cm}
\noindent
\begin{tabular*}{\linewidth}{lc@{\extracolsep{\fill}}r@{\extracolsep{0pt}}}
\vspace*{-1.5cm}\mbox{\!\!\!\includegraphics[width=.14\textwidth]{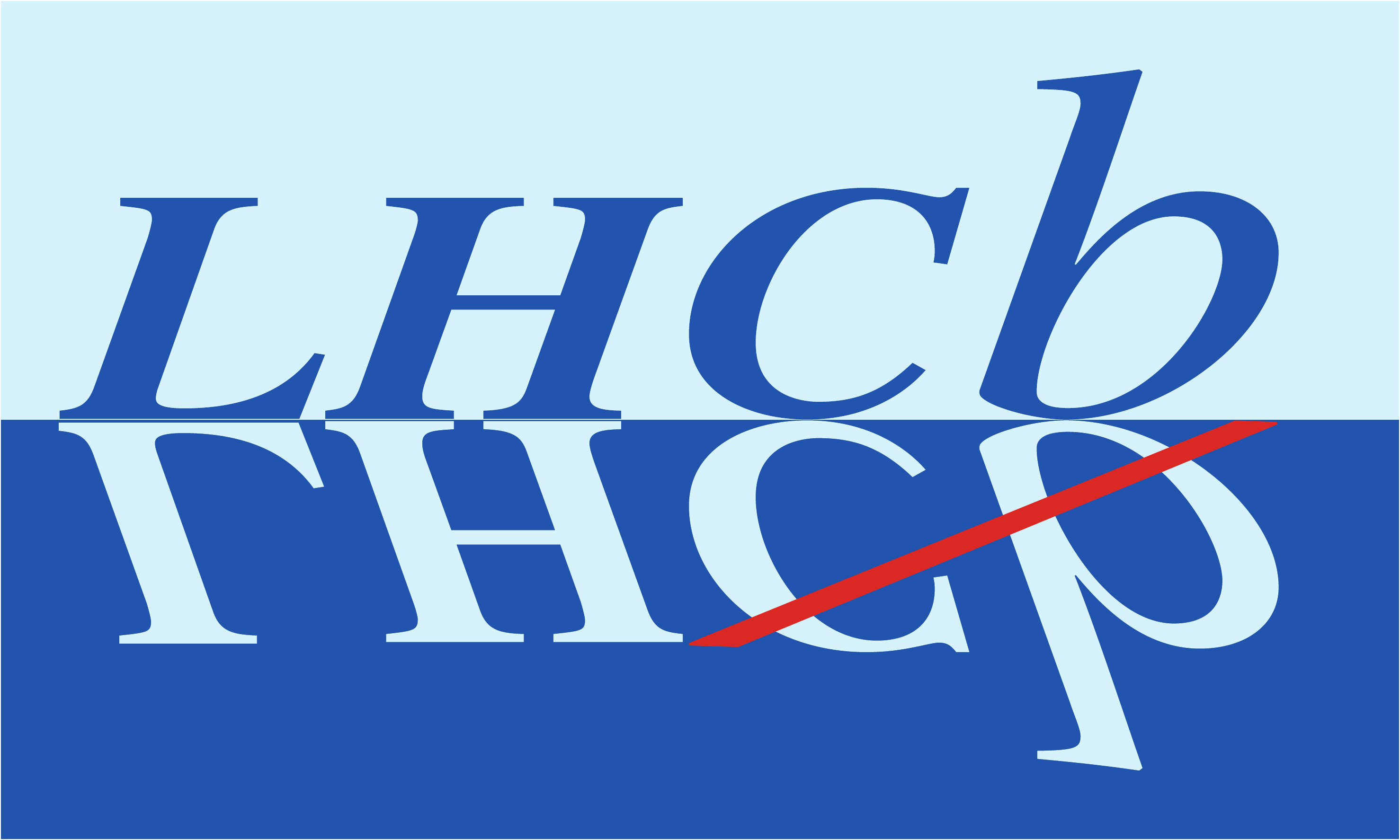}} & &%
\\
 & & CERN-EP-2024-114 \\  
 & & LHCb-PAPER-2024-006 \\  
 & & May 21, 2024 \\ 
 & & \\
\end{tabular*}

\vspace*{4.0cm}

{\normalfont\bfseries\boldmath\huge
\begin{center}
  \papertitle 
\end{center}
}

\vspace*{2.0cm}

\begin{center}
\paperauthors\footnote{Authors are listed at the end of this paper.}
\end{center}

\vspace{\fill}

\begin{abstract}
  \noindent
  A search for the lepton-flavor violating decays $B^0_s\to\phi\mu^\pm\tau^\mp$ is presented, using a sample of proton--proton collisions at center-of-mass energies of 7, 8, and $13\,\text{Te\kern -0.1em V}$, collected with the LHCb detector and corresponding to a total integrated luminosity of $9\,\text{fb}^{-1}$. The $\tau$ leptons are selected using decays with three charged pions. No significant excess is observed, and an upper limit on the branching fraction is determined to be \mbox{$\mathcal{B}\left(B^0_s\to\phi\mu^\pm\tau^\mp\right) < 1.0\times 10^{-5}$} at 90\% confidence level.
  
\end{abstract}

\vspace*{2.0cm}

\begin{center}
  Published in
  Phys.~Rev.~D110 (2024) 7
\end{center}

\vspace{\fill}

{\footnotesize 
\centerline{\copyright~\papercopyright. \href{\paperlicenceurl}{\paperlicence}.}}
\vspace*{2mm}

\end{titlepage}


\newpage
\setcounter{page}{2}
\mbox{~}
%
%
%
%


\renewcommand{\thefootnote}{\arabic{footnote}}
\setcounter{footnote}{0}

\cleardoublepage


\pagestyle{plain} 
\setcounter{page}{1}
\pagenumbering{arabic}


\section{Introduction}
\label{sec:Introduction}

The observation of charged lepton-flavor violation would be an unambiguous sign of physics beyond the Standard Model. Its existence may be implied by hints of lepton flavor nonuniversality in charged-current semileptonic decays of the form \decay{b}{c\ellm\neulb}~\cite{Lees:2013uzd,Huschle:2015rga,Belle:2019rba, LHCb-PAPER-2022-039,LHCb-PAPER-2022-052}, as new physics would likely lead to both phenomena~\cite{Glashow:2014iga}.\footnote{The inclusion of charge-conjugate processes is implied throughout.} Recent theoretical work demonstrates the possibility for relatively large branching fractions for decays involving the \decay{b}{s\mupm\taump} transition~\cite{Bause:2023mfe,He:2023bnk}.

We present a search for the lepton-flavor violating decay \decay{\Bs}{\phi \mu^+ \tau^-},
with  \decay{\phi}{\Kp\Km} and \decay{\taum}{\pim\pip\pim \neut} or \decay{\taum}{\pim\pip\pim \piz \neut}. Since the $\phi$ meson decay does not allow identification of the \Bs flavor, we search for the sum of the four \Bs, \Bsb, $\mup\taum$, and $\mun\taup$ combinations. The two $\taum$ decay signal modes are labeled as \thpi and \thpiz throughout. 
Previous searches at the LHCb, Babar, and Belle experiments have placed upper limits on analogous decays of \Bu and \Bd mesons~\cite{LHCb-PAPER-2022-021,LHCb-PAPER-2019-043,Lees:2012zz,Belle:2022pcr}, and on the related decay \decay{\Bs}{\mup\taum}~\cite{LHCb-PAPER-2019-016}.

We determine the mass of the partially reconstructed signal candidates from a kinematic fit, applying constraints on the \Bs and \taum flight directions and the \taum lepton and unreconstructed neutrino masses.
An extended unbinned maximum-likelihood fit to the constrained-mass distribution determines the signal yield. The signal branching fraction is normalized with respect to the \decay{\Bs}{\psitwos\phi} decay, with \decay{\psitwos}{\jpsi\pip\pim}, \decay{\jpsi}{\mup\mun}, and \decay{\phi}{\Kp\Km}, which has the same final-state particle content as the signal except that a pion is replaced by a muon. An upper limit on the signal branching fraction is set using the Feldman-Cousins method~\cite{Feldman:1997qc}, adapted for the nuisance parameters included in the likelihood~\cite{Bodhisattva:2009uba}.

\section{Detector, data samples, and simulation}
\label{sec:Detector}

The \lhcb detector~\cite{LHCb-DP-2008-001,LHCb-DP-2014-002} is a single-arm forward
spectrometer covering the \mbox{pseudorapidity} range $2<\eta <5$,
designed for the study of particles containing \bquark or \cquark
quarks. The detector includes a high-precision tracking system
consisting of a silicon-strip vertex detector surrounding the $pp$
interaction region, a large-area silicon-strip detector located
upstream of a dipole magnet with a bending power of about
$4{\mathrm{\,Tm}}$, and three stations of silicon-strip detectors and straw
drift tubes placed downstream of the magnet.
The tracking system provides a measurement of the momentum, \ptot, of charged particles with
a relative uncertainty that varies from 0.5\% at low momentum to 1.0\% at 200\gev.\footnote{Natural units with $c = 1$ are used throughout.}
The minimum distance of a track to a primary $pp$ interaction  vertex (PV), the impact parameter,
is measured with a resolution of $(15+29/\pt)\mum$,
where \pt is the component of the momentum transverse to the beam, in\,\gev.
Different types of charged hadrons are distinguished using information
from two ring-imaging Cherenkov detectors. Photons, electrons and hadrons are identified by a calorimeter system consisting of
scintillating-pad and preshower detectors, an electromagnetic
and a hadronic calorimeter.
Muons are identified by a system composed of alternating layers of iron and multiwire proportional chambers. 

The online event selection is performed by a trigger,
which consists of a hardware stage, based on information from the calorimeter and muon
systems, followed by a software stage, which applies a full event
reconstruction.
At the hardware trigger stage, events are required to have a muon with high \pt or a
  hadron, photon or electron with high transverse energy deposited in the calorimeters.
  The software trigger requires a two-, three- or four-track
  secondary vertex with a significant displacement from any primary vertex.

We use data samples collected from 2011 to 2018, at center-of-mass energies of 7, 8, and \SI{13}{\tera\evolt}, corresponding to an integrated luminosity of 9\invfb. We model signal, normalization, and some background decays using simulation. In the simulation, $pp$ collisions are generated using \pythia~\cite{Sjostrand:2006za,*Sjostrand:2007gs} with a specific \lhcb
configuration~\cite{LHCb-PROC-2010-056}. The decays of hadrons and $\taum$ leptons
are described by \evtgen~\cite{Lange:2001uf} and \textsc{TAUOLA}~\cite{tauola}. The signal \bsphimutau decay is described in the simulation using a phase-space decay model. The
interaction of the generated particles with the detector and its response
are implemented using the \geant toolkit~\cite{Allison:2006ve, *Agostinelli:2002hh} as described in Ref.~\cite{LHCb-PROC-2011-006}.

\section{Selection}
\label{sec:selection}

Our \Bs candidate selection begins with a \kkmu combination. The candidates combine tracks consistent with the kaon and muon particle identification (PID) hypotheses and inconsistent with being produced at any PV in the event. The common vertex formed by the three tracks must have a good fit quality and be significantly separated from any PV. We then add three additional tracks with a total charge opposite to that of the muon to form a \Bs candidate---two of these tracks must be consistent with the pion PID hypothesis. In this way we select the signal modes, \thpi and \thpiz, and the normalization mode in which one of the tracks is the other muon from the \decay{\jpsi}{\mup\mun} decay.

We additionally select a control sample for the misidentified background modes \mbox{\decay{\B}{\Dbar \phi \pip}}, including each of the decays \mbox{\decay{\Dzb}{\Kp\pim}}, \mbox{\decay{\Dm}{\Kp\pim\pim}}, and \mbox{\decay{\Dsm}{\Kp\Km\pim}}. Each final-state particle track must be consistent with the appropriate kaon or pion PID hypothesis, and their combined vertex must be of high quality. We combine these charm candidates with two additional kaon candidates and one pion candidate to form the \B candidate. This control sample is discussed further in \cref{sec:background}.

Additional selection criteria are shared by both the signal and normalization modes. The mass of the $\Kp\Km$ pair is required to be within 8\mev of the known $\phi$ meson mass~\cite{PDG2022}; this requirement also applies to the \decay{\B}{\Dbar \phi \pip} control sample. Lower mass thresholds are applied for the \kkmu, $\Kp\Km\pim\pip\pim$, and $\mup\pim\pip\pim$ combinations to remove charm meson contributions. We then separate the normalization mode with high purity by applying mass window requirements for the  \jpsi and \psitwos masses.

\begin{figure}[tb]
\begin{center}
\includegraphics[width=0.7\textwidth]{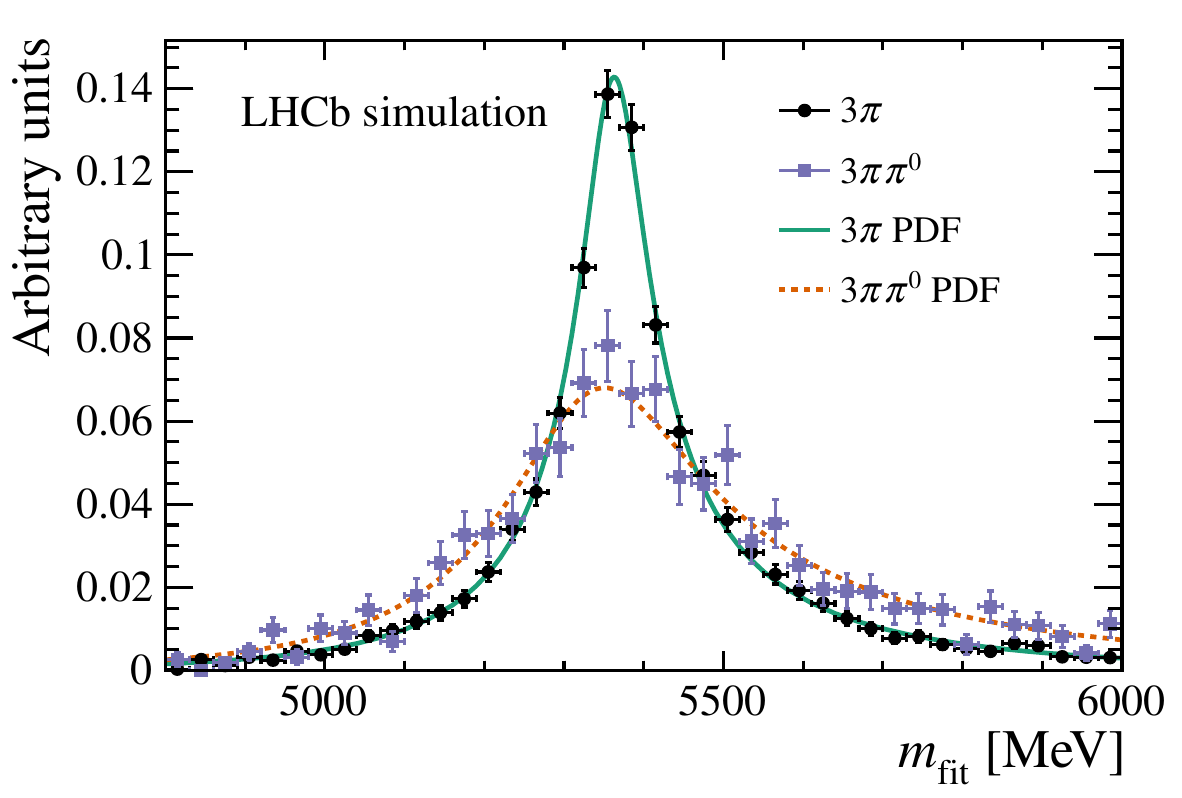}
\end{center}
\caption{Distributions of the constrained mass for simulated \thpi and \thpiz decays. The probability density functions (PDFs) used to model the signal shapes in the final result, are overlaid.\label{fig:sigshapes}}
\end{figure}

Applying direction and mass constraints allows us to reconstruct the mass of the \Bs signal candidates. The directions of the \Bs and $\taum$ candidates are constrained using the PV, the \kkmu secondary vertex, and the \thpi vertex positions. The mass of the $\taum$ lepton is constrained to its known value~\cite{PDG2022} and that of the unreconstructed neutrino to zero. These requirements, which overconstrain the system, are used within a kinematic fit~\cite{Hulsbergen:2005pu}. \Cref{fig:sigshapes} shows the constrained-mass \mfit distribution for simulated \thpi and \thpiz signals. While the \thpiz signal component does not satisfy the zero mass requirement for the unreconstructed $\piz \nu$ pair, the result is only a modest inefficiency and degradation of the mass resolution in this case. The kinematic fit fails to converge in roughly 1\% of cases for signal simulation, and 8\% of cases in the full data sample; in these cases the candidates are rejected from further analysis.

Explicit mass vetoes remove candidates consistent with containing specific decays of charm hadrons, even if the charm decay products were initially separated between the \kkmu and \thpi combinations. We further reduce backgrounds with two Boosted Decision Tree classifiers (BDT)~\cite{Breiman}.  The first is trained to distinguish simulated signal from combinatorial background, which is modeled using wrong-sign candidates in which the \thpi combination has the same total charge as the muon candidate. This BDT relies primarily on topological information, including the goodness of the kinematic fit and the \taum flight distance. It also uses an isolation discriminant that  distinguishes particles from partially reconstructed background decays from other particles in the event that do not originate from the same decay as the signal candidate. A requirement on the BDT output removes the majority of the combinatorial background while being at least 95\% efficient for simulated signal.

The final selection step applies a second BDT classifier that focuses on partially-reconstructed backgrounds from $b$-hadron decays. In the training, this background is modeled with data from the lower \mfit sideband. In addition to topological information about the vertices, the dipion mass distributions for the $\taum$ candidates are important discriminants. We determine the final requirement on this BDT classifier by optimizing the search sensitivity using a simplified version of the final limit procedure in which the background contribution to the signal region is estimated using a linear extrapolation from the lower and upper fitted mass sidebands. 

\section{Normalization}
\label{sec:normalization}

\begin{figure}[tb]
\begin{center}
\includegraphics[width=0.7\textwidth]{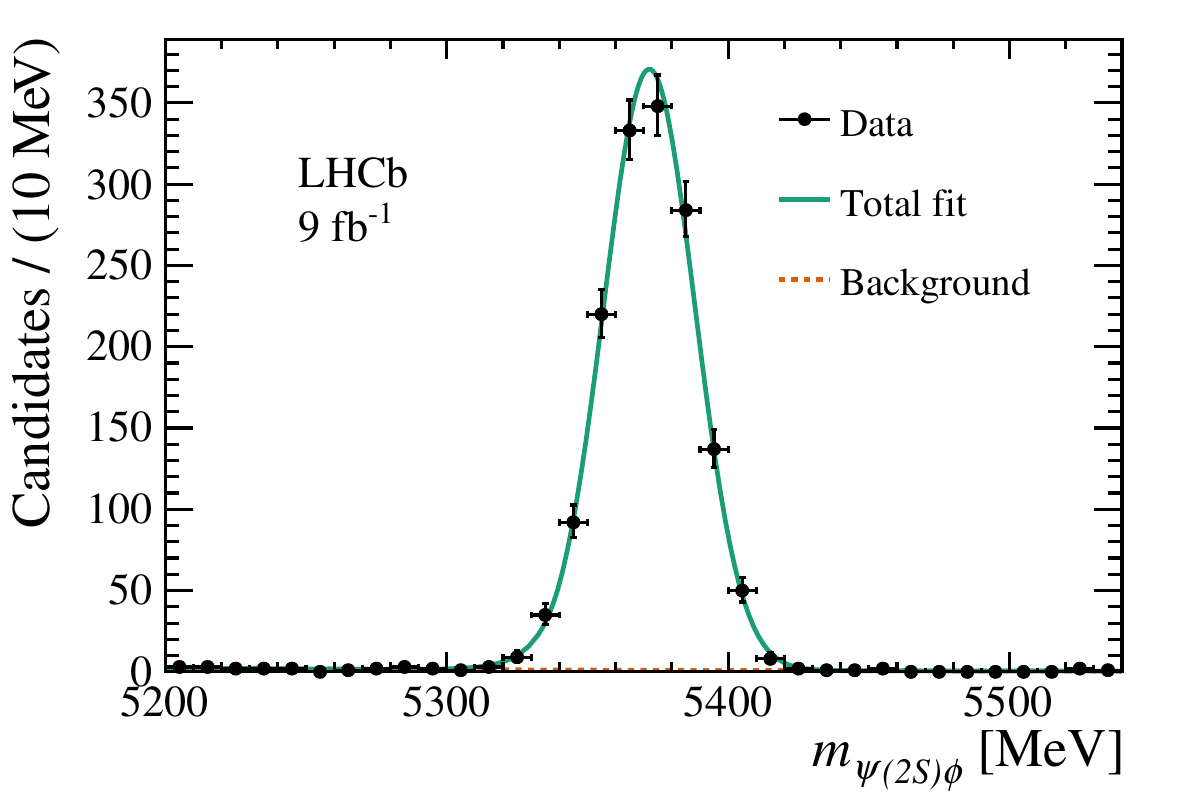}
\caption{Mass distribution of the selected $B^0_s\to\psi(2S)\phi$ decays for the entire data sample. The results of the fit used to determine the shape are overlaid. The normalization mode shape is fixed and used in separate fits per data-taking year to determine the year-by-year normalization yields.\label{fig:norm_fit}}
\end{center}
\end{figure}

We normalize the expected signal yield as a function of its branching fraction, \mbox{$N_{\text{exp}} = \alpha \BF\qty(\decay{\Bs}{\phi\mu\tau})$}, by extrapolating from the normalization-mode yield separately for each data-taking year. Different normalization factors, $\alpha$, are computed for both the \thpi and \thpiz signal modes; for instance
\begin{align}
  \alpha_{3\pi} = &  \frac{  \BF\qty(\decay{\taum}{\pim\pip\pim\neut})}{\BF\qty(\decay{\Bs}{\psitwos\phi})\BF\qty(\decay{\psitwos}{\jpsi\pip\pim})\BF\qty(\decay{\jpsi}{\mup\mun})} \nonumber \\
  & {} \times \sum_{i\in \text{years}} \varepsilon_{\text{rel.},3\pi,i} N_i(\psitwos\phi),
\end{align}
where $\varepsilon_{\text{rel.},3\pi,i}$ is the relative detection efficiency of the \thpi signal mode with respect to the normalization mode in the data-taking year $i$, and $N_i(\psitwos\phi)$ is the observed normalization yield in the same year.
The mass distribution for the normalization channel, combining data-taking years, together with the results of its yield fit, is shown in \cref{fig:norm_fit}. In this fit, the normalization mode is modeled with a Gaussian core with exponential tails on both the upper and lower sides. The small background is modeled with an exponential function. The total yield is 
\mbox{$N\qty(\psitwos\phi) = \num{1490\pm 40}$}
candidates. We fix the shape parameters to the values obtained from this fit in subsequent fits to determine the yields in each subsample.

The ratio of the detection efficiency for the signal to that of the normalization mode is determined using simulation, with several corrections applied to improve the accuracy of the simulation. This includes corrections to the particle identification performance~\cite{LHCb-PUB-2016-021}, the \Bs production kinematics, and the trigger efficiency. These corrections are derived from control samples in data. Before the two BDT selections, the relative efficiencies for the \thpi and \thpiz signal modes are approximately 25\% and 15\%, respectively. After applying the selection criteria on the BDT this falls to 12\% and 5\%. The relative variation of the efficiencies among data-taking years stems primarily from changes in the trigger selection and particle identification performance.

The final normalization factors, including the systematic uncertainties, are
\begin{align*}
  \alpha_{3\pi} &= \num{ 15.3 \pm 1.9 e5 }, \\
  \alpha_{3\pi\piz} &= \num{3.3 \pm 0.4 e5 }.
\end{align*}
The systematic uncertainties are listed in \cref{tab:normsyst}. The largest uncertainty is due to the limited precision of the measured \mbox{\decay{\Bs}{\psitwos\phi}} branching fraction. The uncertainty in the relative efficiency includes contributions from the limited size of the simulation samples, the data-driven corrections, and other sources, including the description of the $\taum$ decay model and the effect of the BDT selection.

The unknown signal decay model results in an uncertainty on the relative efficiency. It is estimated using event-by-event weights in the simulation to match the expected differential branching fraction distributions for different single \decay{b}{s\ell\ell'} effective operator models~\cite{Becirevic:2016zri, Gratrex:2015hna} using form factors from Ref.~\cite{Ball:2004rg}. We do so for decays proceeding via electromagnetic dipole, vector, axial vector, scalar, and pseudoscalar operators. The envelope of the differences obtained from the nominal phase-space model is taken as the uncertainty.

The systematic uncertainties of the $3\pi$ and $3\pi\piz$ signal mode normalization factors are approximately 90\% correlated. They vary in the fit using two Gaussian-constrained nuisance parameters, one for each of the principal components of their covariance.

\begin{table}[tb]
\begin{center}
\caption{Systematic uncertainties on the signal normalization factors for the two signal decay modes.\label{tab:normsyst}}
\begin{tabular}{lSS}
\toprule
Source & \multicolumn{1}{c}{$\delta\alpha_{3\pi}\times 10^{-5}$} & \multicolumn{1}{c}{$\delta\alpha_{3\pi\piz}\times 10^{-5}$} \\
\midrule
$\psitwos\phi$ branching fraction                        & 1.18 & 0.26 \\ 
Other branching fractions                                 & 0.18 & 0.05 \\ 
Normalization yield                              & 0.40 & 0.09 \\ 
Size of simulated samples         & 0.47 & 0.15 \\ 
$\tau$ decay model                        & 0.61 & 0.13 \\ 
Signal decay model                             & 0.90 & 0.18 \\ 
Other efficiency uncertainties & 0.68 & 0.16 \\
\midrule 
Total                         & 1.9 & 0.4 \\ 
\bottomrule
\end{tabular}
\end{center}
\end{table}

\section{Background estimation}
\label{sec:background}

We expect that the majority of the remaining background after selection is a sum of contributions from combinatorial backgrounds and a number of different partially reconstructed $b$-hadron decays, many of which include two intermediate charm mesons. These decays do not have the correct topology to satisfy the direction and mass constraints applied in the kinematic fit, and therefore vary smoothly across the mass distribution in the final fit. This expectation is based on estimations of the expected yield and mass distribution under analytic mass reconstruction techniques from fast simulations for many different decay modes with two charm mesons~\cite{Cowan:2016tnm}.
 Fits to a number of control samples validate this expectation; they are selected from the $\phi$ mass sidebands, the wrong-sign $\mup\taup$ sample, and a sample enriched with misidentified muons. Otherwise, these samples pass all signal selection requirements. These studies motivate the use of the smooth background models described in \cref{sec:fit}.

However, we do expect a non-negligible contribution from decays of the form \mbox{\decay{\B}{\Dbar \Kp\Km \pip}} in which the $\pip$ is misidentified as a muon. The expected yield for these decays is estimated using simulation and control samples. Because of the similarity to the signal mode, these decays peak in the reconstructed \mfit and pass the multivariate selections with relatively high efficiency.

The size of this background contribution is estimated using the fully reconstructed $\Dbar$ decays listed in \cref{sec:selection}. We determine the yield for \Bd, \Bu, and \Bs decays to these final states using fits to the reconstructed \B mass distributions. The resulting control yields and the extrapolated contributions to the signal sample are listed in \cref{tab:dkkpi_extrap}. Two sets of simulated samples are used to extrapolate these yields to the final signal sample. The first set includes decays that match the control sample in data, and the second includes a sum of many exclusive decays of the form \decay{\Dbar}{\pim\pip\pim X}, where $X$ refers to any number of unreconstructed particles. The efficiency for pions to pass muon particle identification requirements is estimated using a calibration sample in data. A large fraction of pions passing these requirements decay to muons in the detector volume, and may also pass the muon trigger requirements. The muon trigger efficiency calibration used for the signal is taken as an upper bound on the efficiency for this background. We assign a systematic uncertainty for the misidentification rate of 20\%, based on studies using the calibration samples.

The inclusive \decay{\Dbar}{\pim\pip\pim X} decay rates have been measured by BESIII~\cite{BESIII:2023sxm,BESIII:2022ydh}. However, the composition of \decay{\Dbar}{\pim\pip\pim X} decays is uncertain, and the efficiency to pass the signal selection is sensitive, for example, to the rates of decays with intermediate \rhoz mesons. The simulated samples use a number of previously measured exclusive decay rates~\cite{PDG2022}, extrapolating, for example, for decays with \KL instead of \KS mesons. The sum of these exclusive rates underestimates the inclusive rates measured by BESIII, but includes the decays most similar to the signal, which are most likely to be selected. To account for this, we assign a large systematic uncertainty on this extrapolation equal to the total difference between the measured rates and the sum-of-exclusives rates for each of the \Dzb, \Dm, and \Dsm mesons, which are 40\%, 32\%, and 47\% respectively.
The final yield of this background is estimated to be  \num{0.63 \pm 0.25} decays with the dominant uncertainty coming from the \decay{\Dbar}{\pim\pip\pim X} extrapolation.

\begin{table}[tb]
\caption{Measured sum of $B^0$, $B^+$, and $B^0_s$ decays of the form $\kern 0.18em \overline{\kern -0.18em D} K^+ K^-\pi^+$ for each of the three studied charm mesons. Using simulation and calibration samples, these control yields are extrapolated to the expected yield in the final signal selection with the pion misidentified as a muon. The uncertainties include only the statistical uncertainty from the control yield fits.\label{tab:dkkpi_extrap}}
\begin{center}
\begin{tabular}{cS[table-format = 3.0(2)]S[table-format = 1.2(1)]}
\toprule
Charm meson & \multicolumn{1}{c}{Control yield} & \multicolumn{1}{c}{Expected misid.} \\
\midrule
\Dm & 309 \pm 31 & 0.11 \pm 0.01 \\ 
\Dzb & 343 \pm 27 & 0.07 \pm 0.01 \\ 
\Dsm & 632 \pm 39 & 0.45 \pm 0.03 \\
\bottomrule
\end{tabular}
\end{center}
\end{table}

\section{Fit and limit setting procedure}
\label{sec:fit}

We perform an extended unbinned maximum-likelihood fit of the \mfit distribution. The parameter of interest is the signal branching fraction. The two signal shapes are determined from simulation; they are described by generalized hyperbolic functions~\cite{sigshape}, which are shown in \cref{fig:sigshapes}. The shape parameters vary in the fit; they are constrained to the results of the fit to the simulation with Gaussian constraints along the principal components of the covariance matrices. We include in the same way the peaking, misidentified backgrounds; one shape describes the sum of their contributions. The remaining background is modeled using four background parameterizations: exponential, linear, quadratic, and the product of an exponential and linear function. Each background parameter is unconstrained, and each background function is restricted to positive values over the fit range. We treat the choice of the background function as a discrete nuisance parameter~\cite{Dauncey:2014xga}, applying a penalty of one unit to twice the negative logarithm of the likelihood function for the two background models with two free parameters. We additionally require the signal branching fraction to be non-negative; this corresponds to using the test statistic
\begin{equation}
  \tilde{t}_\BF = \begin{cases}
    -2\log\qty( \frac{ \mathcal{L}\qty(\BF, \doublehat{\theta}\qty(\BF)) }{ \mathcal{L}\qty(\hat{\BF},\hat{\theta}) } ) & \hat{\BF} \geq 0 \\
    -2\log\qty( \frac{ \mathcal{L}\qty(\BF, \doublehat{\theta}\qty(\BF)) }{ \mathcal{L}\qty(0,\doublehat{\theta}\qty(0)) } ) & \hat{\BF} < 0,
  \end{cases}
\end{equation}
where $\mathcal{L}$ is the likelihood including the penalty, $\BF$ the signal branching fraction, $\theta$ the vector of nuisance parameters, $\hat{\BF}$ and $\hat{\theta}$ the best-fit parameters, and $\doublehat{\theta}(\BF)$ the best-fit nuisance parameters when fixing $\BF$ to a particular value. 

The sampling distributions for the test statistic $\tilde{t}_\BF$ are obtained for different values of \BF using pseudoexperiments. For a particular value of \BF, we use the conditional best-fit background description with fixed signal branching fraction to generate the ensemble. We compute the $p$-value as a function of \BF by integrating the sampling distributions above the observed test statistic.

\section{Results and conclusion}
\label{sec:results}

\begin{figure}[tbp]
\includegraphics[width=0.5\textwidth]{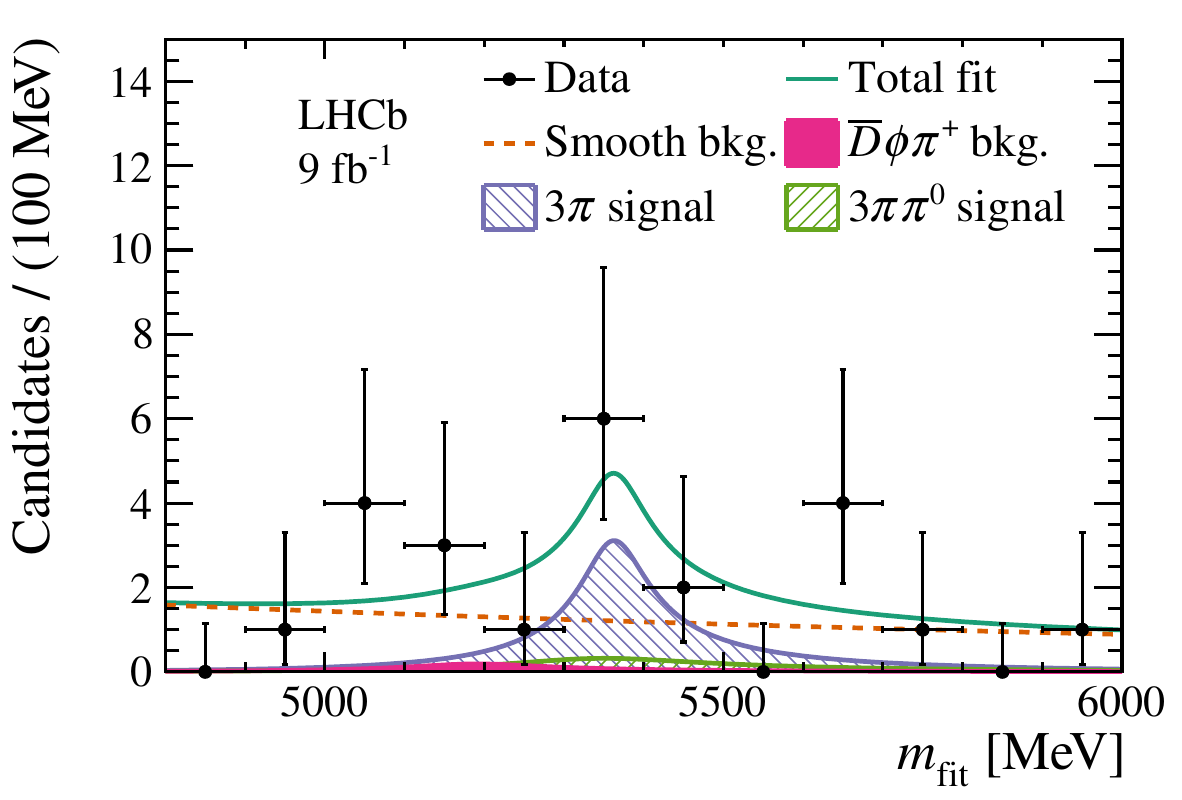}
\includegraphics[width=0.5\textwidth]{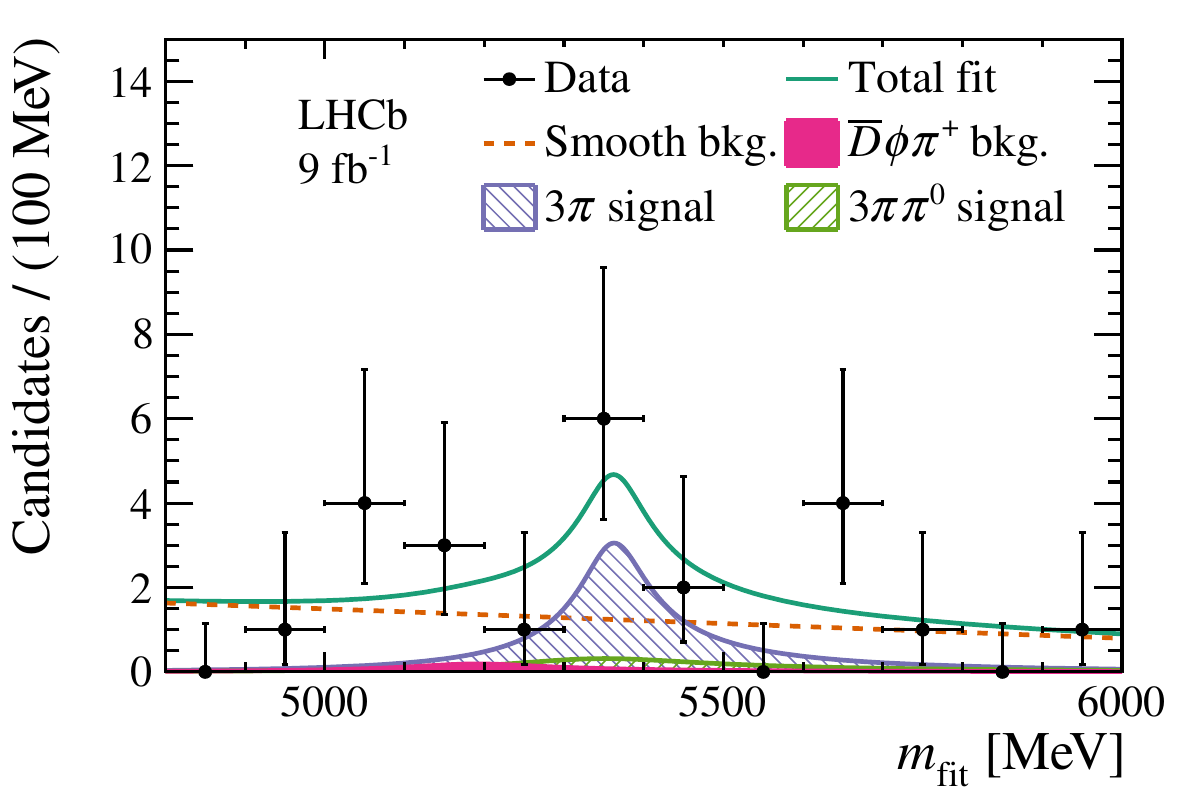} \\
\includegraphics[width=0.5\textwidth]{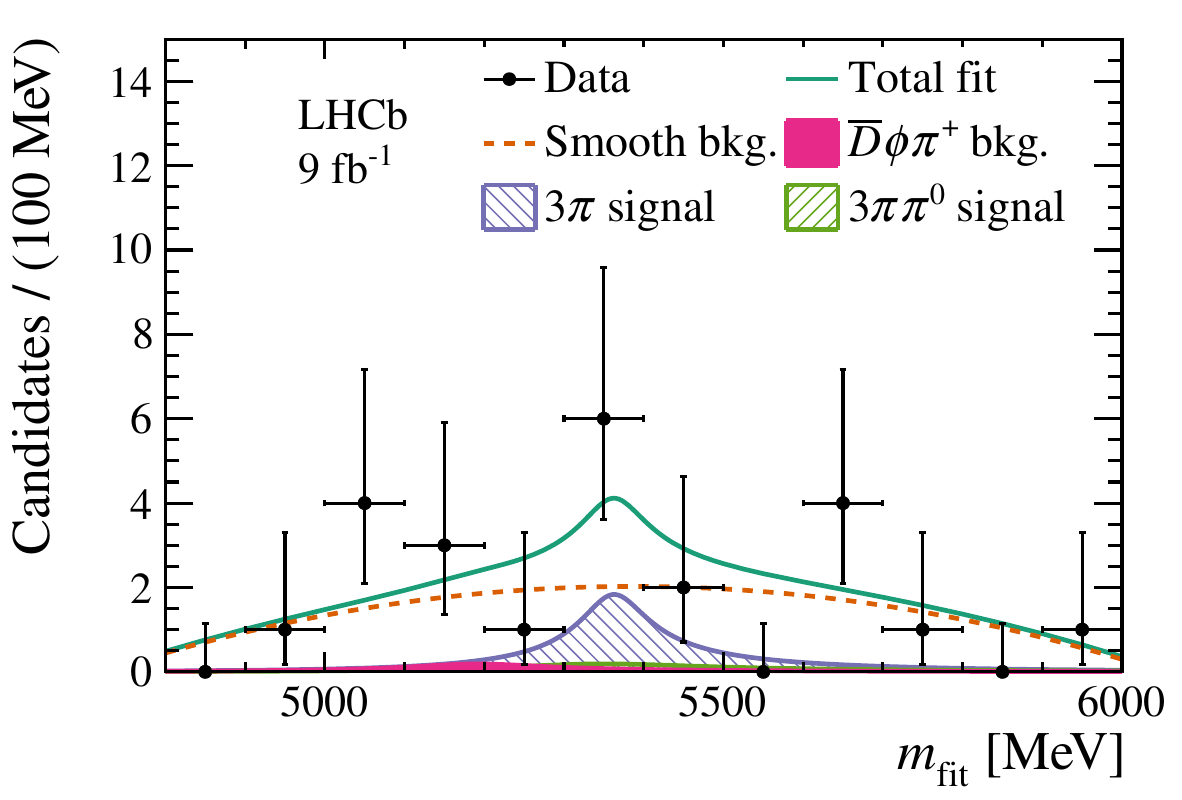}
\includegraphics[width=0.5\textwidth]{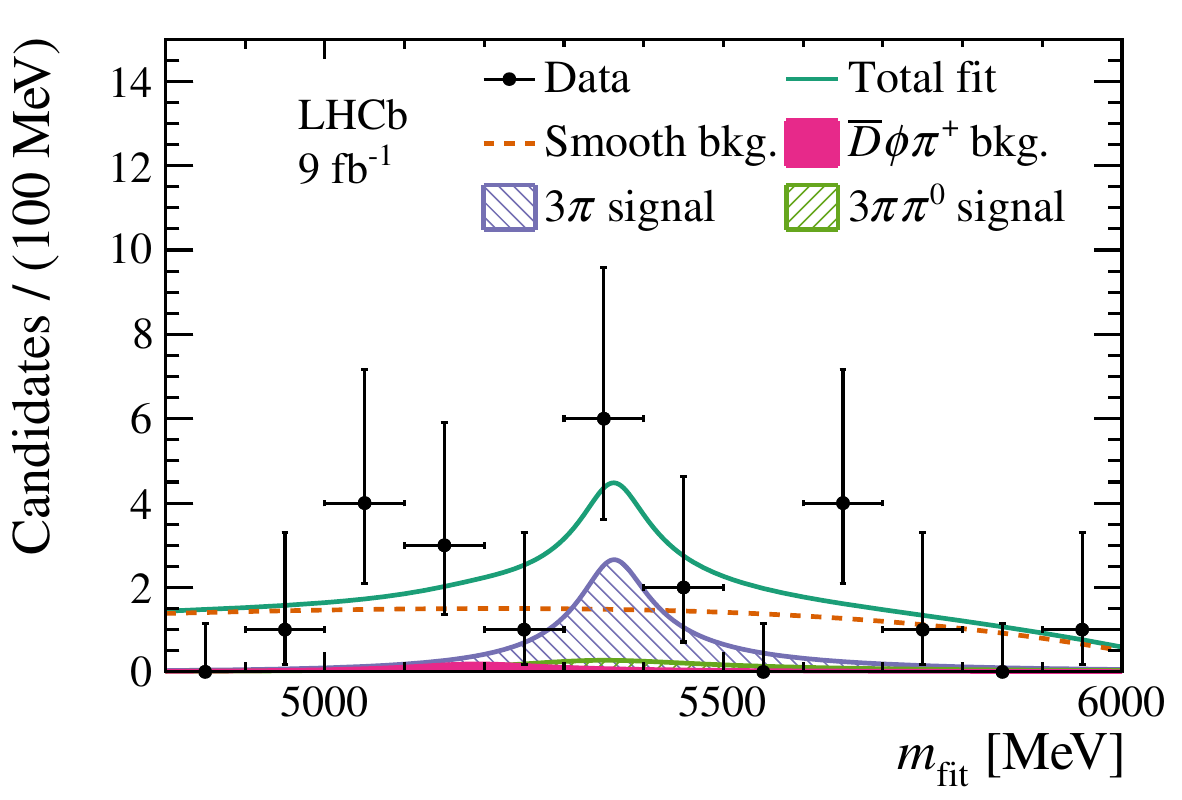}
\caption{Distributions of $m_{\text{fit}}$ overlaid with the fit results corresponding to four background models: (top left) exponential, (top right) linear, (bottom left) quadratic, and (bottom right) the product of linear and  exponential functions. The linear background shape provides the overall best-fit point. The peaking $\kern 0.18em \overline{\kern -0.18em D}\phi\pi^+$ background with a misidentified muon and the two signal shapes are also shown.\label{fig:bestfit}}
\end{figure}

\begin{figure}[tbp]
\includegraphics[width=0.5\textwidth]{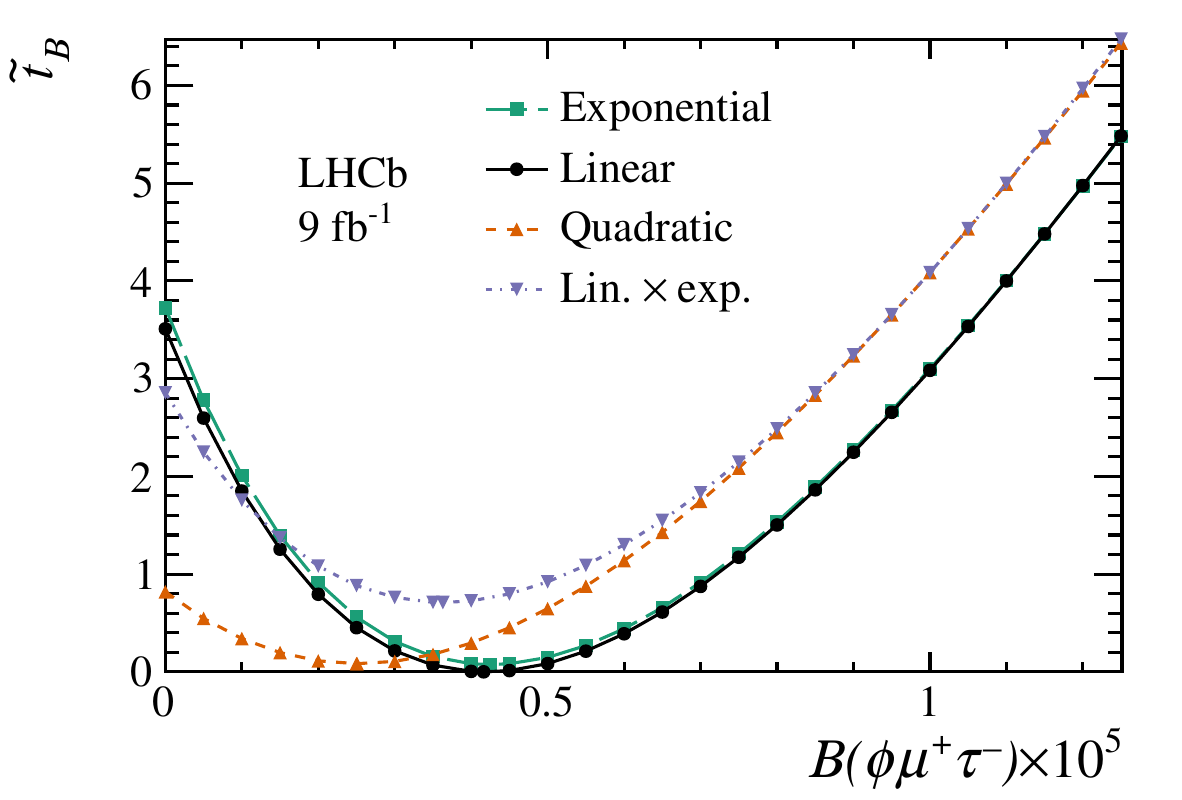}
\includegraphics[width=0.5\textwidth]{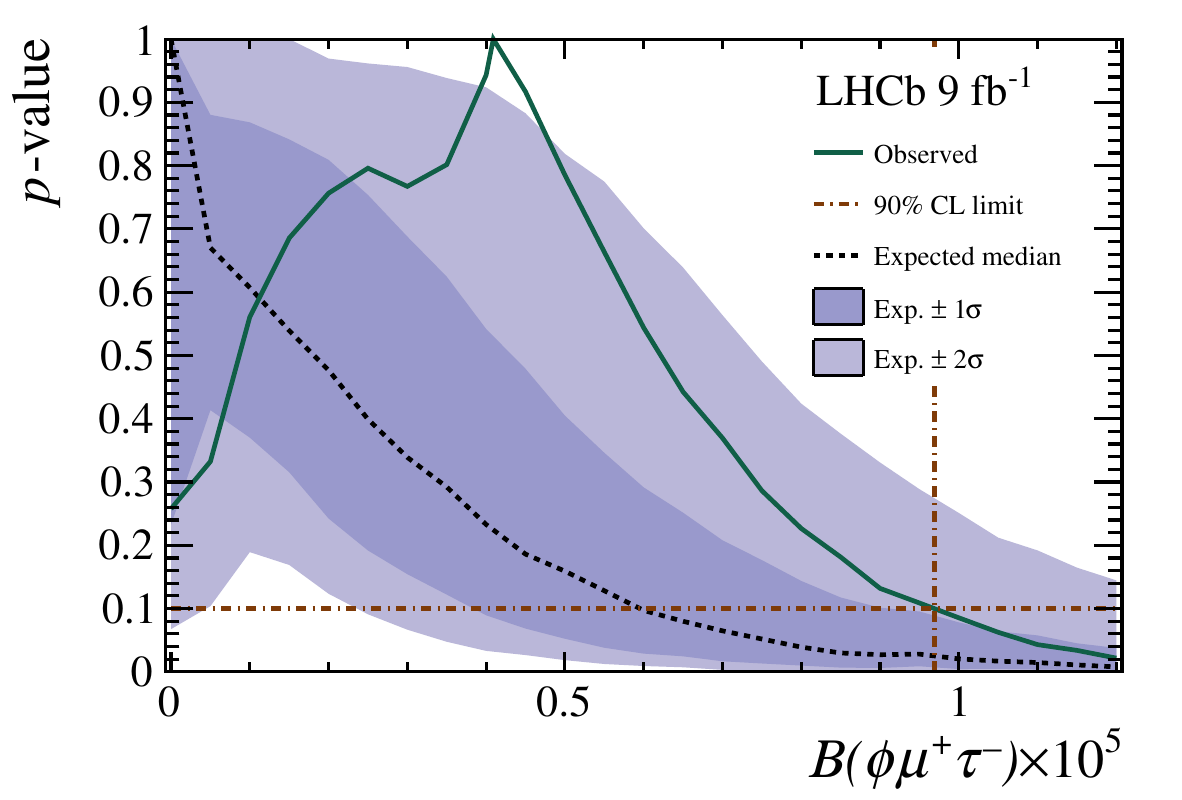}
\caption{Left: Observed test-statistic scans for each of the four background shapes (exponential, linear, quadratic, and the product of linear and exponential functions). Right: the $p$-values obtained for the signal branching fraction from the observed test-statistics and their sampling distributions, together with the median expected limit and its $1\sigma$ and $2\sigma$ intervals. The limit at 90\% CL is indicated by the crossed lines.\label{fig:nllscan}}
\end{figure}

The best-fit results for each of the four background models are shown in \cref{fig:bestfit}. The overall best fit uses the linear background model, and results in a signal branching fraction of \num{4.1e-6}. The observed test-statistic scans for each background model, including the penalty term, are shown in \cref{fig:nllscan} (left); for a particular value of \BF, the lowest value among these four test statistics is the value $\tilde{t}_\BF$.

The upper limit for \BF is derived from the $p$-value curve shown in \cref{fig:nllscan} (right). The result is 
\begin{align*}
    \BF(\bsphimutau) &< \num{1.0e-5} \text{ at 90\% CL,} \\
    \BF(\bsphimutau) &< \num{1.1e-5} \text{ at 95\% CL.}
\end{align*}
This is the first limit on this lepton-flavor violating decay. The sensitivity is competitive with similar searches for \mup\taum pairs in \Bu and \Bd decays.

\section*{Acknowledgements}
%
%
\noindent We express our gratitude to our colleagues in the CERN
accelerator departments for the excellent performance of the LHC. We
thank the technical and administrative staff at the LHCb
institutes.
We acknowledge support from CERN and from the national agencies:
CAPES, CNPq, FAPERJ and FINEP (Brazil); 
MOST and NSFC (China); 
CNRS/IN2P3 (France); 
BMBF, DFG and MPG (Germany); 
INFN (Italy); 
NWO (Netherlands); 
MNiSW and NCN (Poland); 
MCID/IFA (Romania); 
MICIU and AEI (Spain);
SNSF and SER (Switzerland); 
NASU (Ukraine); 
STFC (United Kingdom); 
DOE NP and NSF (USA).
We acknowledge the computing resources that are provided by CERN, IN2P3
(France), KIT and DESY (Germany), INFN (Italy), SURF (Netherlands),
PIC (Spain), GridPP (United Kingdom), 
CSCS (Switzerland), IFIN-HH (Romania), CBPF (Brazil),
and Polish WLCG (Poland).
We are indebted to the communities behind the multiple open-source
software packages on which we depend.
Individual groups or members have received support from
ARC and ARDC (Australia);
Key Research Program of Frontier Sciences of CAS, CAS PIFI, CAS CCEPP, 
Fundamental Research Funds for the Central Universities, 
and Sci. \& Tech. Program of Guangzhou (China);
Minciencias (Colombia);
EPLANET, Marie Sk\l{}odowska-Curie Actions, ERC and NextGenerationEU (European Union);
A*MIDEX, ANR, IPhU and Labex P2IO, and R\'{e}gion Auvergne-Rh\^{o}ne-Alpes (France);
AvH Foundation (Germany);
ICSC (Italy); 
Severo Ochoa and Mar\'ia de Maeztu Units of Excellence, GVA, XuntaGal, GENCAT, InTalent-Inditex and Prog. ~Atracci\'on Talento CM (Spain);
SRC (Sweden);
the Leverhulme Trust, the Royal Society
 and UKRI (United Kingdom).

\addcontentsline{toc}{section}{References}
\bibliographystyle{LHCb}
\bibliography{main,standard,LHCb-PAPER,LHCb-CONF,LHCb-DP,LHCb-TDR}

\newpage
\centerline
{\large\bf LHCb collaboration}
\begin
{flushleft}
\small
R.~Aaij$^{36}$\lhcborcid{0000-0003-0533-1952},
A.S.W.~Abdelmotteleb$^{55}$\lhcborcid{0000-0001-7905-0542},
C.~Abellan~Beteta$^{49}$,
F.~Abudin{\'e}n$^{55}$\lhcborcid{0000-0002-6737-3528},
T.~Ackernley$^{59}$\lhcborcid{0000-0002-5951-3498},
A. A. ~Adefisoye$^{67}$\lhcborcid{0000-0003-2448-1550},
B.~Adeva$^{45}$\lhcborcid{0000-0001-9756-3712},
M.~Adinolfi$^{53}$\lhcborcid{0000-0002-1326-1264},
P.~Adlarson$^{79}$\lhcborcid{0000-0001-6280-3851},
C.~Agapopoulou$^{13}$\lhcborcid{0000-0002-2368-0147}
% ...

{\footnotesize \it

% ...
$^{13}$Universit{\'e} Paris-Saclay, CNRS/IN2P3, IJCLab, Orsay, France\\
% ...
$^{36}$Nikhef National Institute for Subatomic Physics, Amsterdam, Netherlands\\
% ...
$^{45}$Instituto Galego de F{\'\i}sica de Altas Enerx{\'\i}as (IGFAE), Universidade de Santiago de Compostela, Santiago de Compostela, Spain\\
% ...
$^{49}$Physik-Institut, Universit{\"a}t Z{\"u}rich, Z{\"u}rich, Switzerland\\
% ...
$^{53}$H.H. Wills Physics Laboratory, University of Bristol, Bristol, United Kingdom\\
% ...
$^{55}$Department of Physics, University of Warwick, Coventry, United Kingdom\\
% ...
$^{58}$School of Physics and Astronomy, University of Glasgow, Glasgow, United Kingdom\\
$^{59}$Oliver Lodge Laboratory, University of Liverpool, Liverpool, United Kingdom\\
% ...
$^{67}$Syracuse University, Syracuse, NY, United States\\
% ...
$^{79}$Department of Physics and Astronomy, Uppsala University, Uppsala, Sweden, associated to $^{58}$\\
% ...
}
\end{flushleft}

\end{document}